# Seq2Bind Webserver for Decoding Binding Hotspots directly from Sequences using Fine-Tuned Protein Language Models

Xiang Ma[1,2,†], Supantha Dey[3, †], Vaishnavey SR[3],
Casey Zelinski[3], Qi Li[1, *], and Ratul Chowdhury[3, *]

[1] Department of Computer Science, Iowa State University, Ames, Iowa, USA 50011
[2] Department of Chemistry, Grand View University, Des Moines, Iowa, USA 50316
[3] Department of Chemical and Biological Engineering, Iowa State University, Ames, Iowa, USA 50011

† Joint Authors
* To whom correspondence should be addressed. Email: ratul@iastate.edu
* Co-corresponding author: Qi Li

**GRAPHICAL ABSTRACT**

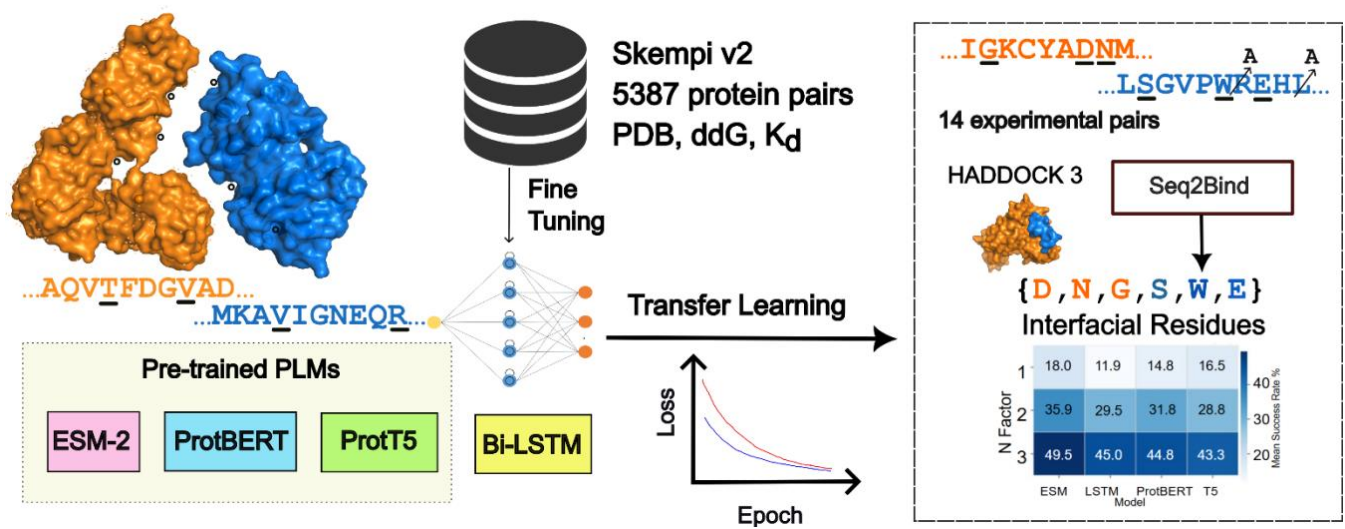

**ABSTRACT**

Decoding protein-protein interactions (PPIs) at the residue level is crucial for understanding cellular mechanisms and developing targeted therapeutics. We present Seq2Bind Webserver, a computational framework that leverages fine-tuned protein language models (PLMs) to determine binding affinity between proteins and identify critical binding residues directly from sequences, eliminating the structural requirements that limit most affinity prediction tools. We fine-tuned four architectures including ProtBERT, ProtT5, ESM2, and BiLSTM on the SKEMPI 2.0 dataset containing 5,387 protein pairs with experimental binding affinities. Through systematic alanine mutagenesis on each residue for 14 therapeutically relevant protein complexes, we evaluated each model's ability to identify interface residues. Performance was assessed using N-factor metrics, where N-factor=3 evaluates whether true residues appear within 3n top predictions for n interface residues. ESM2 achieved 49.5% accuracy at N-factor=3, with both ESM2 (37.2%) and ProtBERT (35.1%) outperforming structural docking method HADDOCK3 (32.1%) at N-factor=2. Our sequence-based approach enables rapid screening (minutes versus hours for docking), handles disordered proteins, and provides comparable accuracy, making Seq2Bind a valuable prior to steer blind docking protocols to identify putative binding residues from each protein for therapeutic targets. Seq2Bind Webserver is accessible at https://agrivax.onrender.com under StructF suite.

**Keywords**: protein-protein interaction, protein language model, binding affinity, mutation scanning, sequence-based model.

## INTRODUCTION

Protein-protein interactions (PPIs) play a pivotal role in facilitating various cellular processes, influencing the regulation of gene expression, signal transduction, and the formation of complex biological networks.(1–3) The intricate interplay between proteins affects essential biological functions, and deciphering the landscape of PPIs is crucial for understanding cellular mechanisms and disease pathways.(4–7) Traditional experiments (8,9) and high-throughput techniques (10,11) have generated a significant amount of PPI data, but they involve expensive and time-consuming wet-lab experiments. Accurate prediction of PPIs by machine learning and deep learning is, therefore, of paramount importance in expediting campaigns for comprehension of cellular functions and drug discovery efforts. The ability to predict PPI strengths holds significant promise for unveiling the intricacies of cellular interactions (such as signal transduction, competitive inhibition, and inter-cell quorum sensing) and guiding targeted therapeutic interventions.

Traditional machine learning methods, such as Support Vector Machine (SVM) (12) and Random Forest (RF) (13), have played a foundational role in developing effective PPI predictors. Given the primary influence of protein sequences on interactions, much research has concentrated on sequence-based PPI prediction. The key step in constructing a machine learning model for PPI involves feature engineering, converting the protein sequences into fixed-dimensional vectors. Amino Acid Composition (AAC), Dipeptide Composition (DPC), Conjoint Triad (CT), Doc2Vec and Position-specific Scoring Matrix (PSSM) are prevalent sequence encoding schemes (14–18), providing descriptive representations and are often leveraged by traditional machine learning methods.(19,20) This harnesses the inherent information within protein sequences, offering a robust strategy for deciphering complex protein-protein interaction strength, and structural/ biochemical priors that drive these interactions. However, traditional machine learning methods often face limitations, such as nonlinear relationships and feature extraction, in capturing intricate and nuanced relationships within highly dynamic and complex biological systems. (21)

Recent advancements in deep learning, particularly transfer learning approaches borrowed from the natural language processing (NLP) domain, have demonstrated significant advantages in predicting PPIs compared to traditional machine learning methods. Notably, attention-based models, including Bidirectional Encoder Representations from Transformers (BERT) and Text-To-Text Transfer Transformer (T5), have been adapted from NLP and trained on a large corpus of protein sequences in a self-supervised fashion to learn biophysical features inherent in protein sequences.(22) In this context, protein language models (PLMs) treat sequences as analogous to sentences, with amino acids serving as the fundamental building blocks, akin to a natural vocabulary. These pre-trained PLMs have shown to serve as promising foundational bases for downstream tasks to map sequence to function. While there has been a great deal of focus on leveraging protein language models for accurate prediction of 3D protein structures individually (RGN2 (23), AlphaFold2(24), OpenFold (25), ESMFold (26), OmegaFold (27)) or in complex with other proteins (AlphaFold-Multimer (28)) - the inference times for structure prediction is long and is also beset with uncertainties proportional to the number of known family members for a protein. In addition, protein-protein docking protocols also put forward uncertainties on binding poses depending on how structured and where the binding loci are. This is why predicting antigen interactions with hypermutable, unstructured CDR3 stretches of antibodies are still elusive to structure-based methods. Protein language-model encodings (an implicit summary of both sequence and structure) is gaining traction for protein function prediction, enzyme activity (29), distant homology detection (30), allosteric site prediction (31), subcellular localization (32) etc. However, training a binding affinity predictor between two proteins, and using it downstream to directly predict the hotspot amino acids that are key for the interaction between proteins has thus far remained mostly unaddressed.

In this study, we addressed this gap and have put forward a general computational assessment platform (Seq2Bind) to identify promising pre-trained natural language encodings of proteins and fine-tuned them for protein pairs and jointly trained them to predict normalized experimental binding strength between them. We fine-tuned four different pre-trained architectures and subsequently queried the trained models to pinpoint key amino acids from each protein in a pair which when altered compromises the binding strength – i.e., critical to binding. We iteratively altered amino acid types along the polypeptide backbones of both proteins in a pair to alanine (one at a time) to glean those residue positions which lead to highest drops in binding energy. Experimentally confirmed crystal structures from 14 unseen protein pairs (relevant for human health), were used for a blind test to see if the models can truly identify interacting residues at the interface. These serve as exemplars for our validation approach. Subsequently, residues (and some of their neighboring flanks) from both proteins were all substituted by alanine resulted in augmented signals i.e., reveal with certainty whether the language model can capture a locus' importance in mediating interaction with the partner protein.

In parallel, we also took these known 14 protein pairs from the Protein Data Bank and split their chains and re-docked them using structural docking program HADDOCK3 to compute the number of native interaction residues that the top HADDOCK3 pose recovers. For the set of 14 protein pairs fine-tuned best Seq2Bind model shows 18.4% accuracy when compared to HADDOCK3. By presenting an in-depth analysis of the predictive performance of PLMs on mutated protein sequences, we show that subtle point mutations are difficult to be translated into local changes in protein geometry that could (a) be picked up by deep learning-based protein structure predictors, and (b) show any difference in structure-based docking protocols. The mutational landscape poses a unique challenge for predictive models, and assessing their robustness in predicting interacting residues under varying mutational scenarios is crucial for establishing their reliability. This study offers insights of protein language models in its adaptability to mutated protein sequences for not only structured proteins but also non-structural transmembrane viral proteins which might be important for entry into a host (with consequences in disease surveillance, and antibody design for intervention in humans, animals, and plants). Our results shed new light on leveraging the *in-situ* potential of PLMs for predicting key residues that biochemically drive protein-protein interaction.

## METHODS

### Datasets

The protein dataset utilized in this study was retrieved from the publicly accessible Structural Kinetics of Energetic and Molecular Processes Interface (SKEMPI 2.0) database (33). The data set, encompassing a wide range of protein complexes, including endogenous and exogenous interactions, comprises 7086 pairs of proteins with experimentally determined thermodynamic data. Utilizing the affinity data, represented by the dissociation constant ($K_D$), we computed the negative delta G as the target affinity score using the formula

$$-\Delta G = 298 \times 1.9872 \times \frac{\ln K_D}{1000} \quad (1)$$

Protein sequences were extracted from the provided PDB files, and mutations were applied using mutation codes (e.g., LI38G denotes replacing L at position 38 in chain I with G). The accuracy of amino acid residues was verified against the protein sequences. We retrieved 344 wild-type protein pairs. Subsequently, 287 pairs lacking affinity data and 1573 pairs with mismatches between mutation codes and protein sequences were excluded. After removing 183 duplicated data points, we obtained 5387 pairs of proteins for analysis. This processed dataset was then divided into two subsets: 70% (3770 data points) for training, and 30% (1617 data points) for validation purposes.

**Models**

We employed three PLMs, ProtBERT, ProtT5, (22) and Evolutionary Scale Modeling 2 (ESM2) (34), for the embedding of protein sequences. ProtBERT and ProtT5 are based on the BERT and T5 models, both of which are deep learning architectures recently developed in the NLP field. ESM2, based on transformer architecture, has become a cornerstone in modern NLP and computational biology due to its ability to model long-range dependencies in sequences. The PLMs underwent a pretraining process on large-scale datasets using a masked language modeling (MLM) objective. Specifically, during training, a masked language model randomly masks 15% of the amino acids in the input sequences, fostering an efficient learning process. The ProtBERT model was pretrained on the Uniref100 dataset, comprising a 217 million protein sequences, while the ProtT5-XL-UniRef50 model, based on the t5-3b architecture, and ESM2 (esm2_t33_650M_UR50D) were pretrained on UniRef50, a dataset containing 45 million protein sequences. (22) The encoder portion (prot_t5_xl_half_uniref50-enc) of the original ProtT5-XL-UniRef50 model was used in this study.

The embedding process of Seq2Bind begins with tokenization of a given protein sequence, followed by the addition of positional encoding to each token. This tokenized input is then passed through stacked self-attention layers, enabling the generation of context-aware embeddings. Each residue within the protein sequence is encoded into an output embedding of length 1024. The hidden state of the attention stack, with dimensions corresponding to the number of amino acids, is utilized as a feature matrix for downstream tasks, e.g., regression or classification.

PLMs were integrated into a Siamese neural network architecture (Figure 1), a design inspired and modified from recent literature (35). The regression head of the network was modified to include three dropouts and linear layers. The final linear layer was configured to have one output, predicting the binding energy. Additionally, a ReLU activation function was applied to the output layer to ensure that all scores were positive. Various loss functions, including mean squared error (MSE), mean absolute error (MAE), and Huber, were tested, along with different batch sizes, learning rates, and weight decay hyper-parameters, to optimize performance. The MSE and MAE metrics were used as evaluative measures to assess the model's performance.

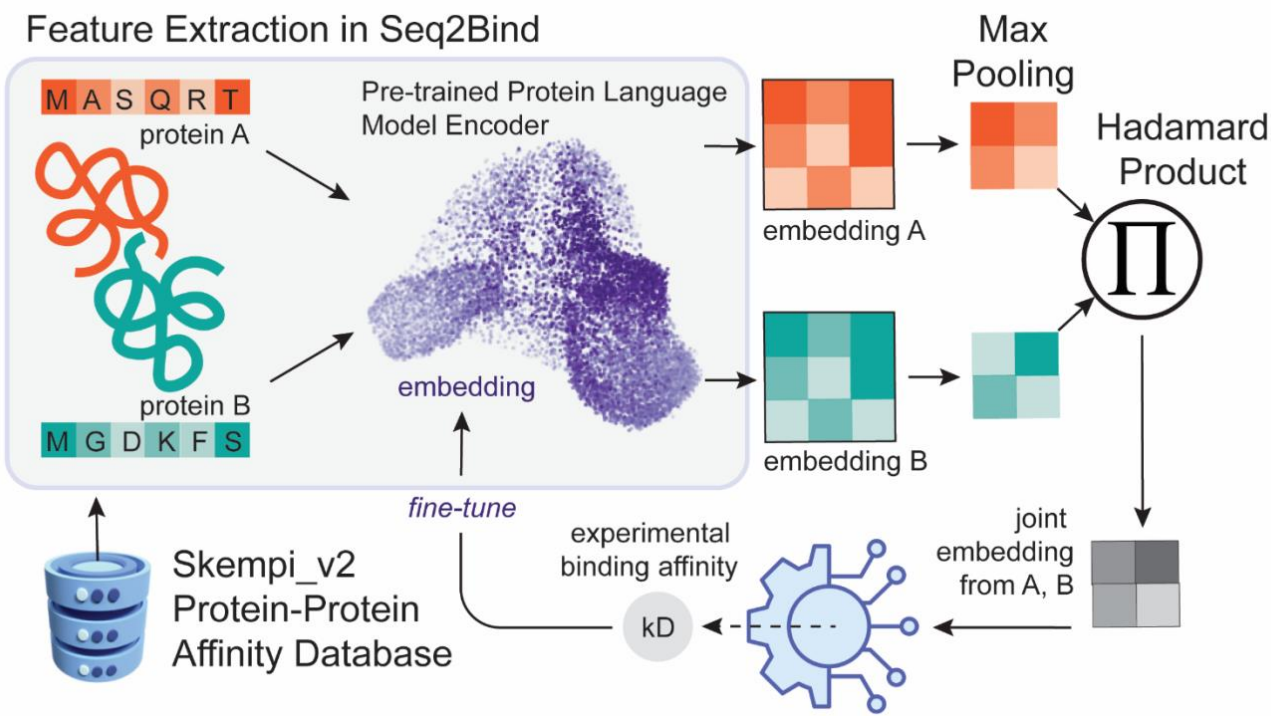

**Figure 1.** Schematic representation of the deep learning workflow for predicting experimental protein-protein binding affinity. Two protein sequences are tokenized and processed through pre-trained PLMs to generate sequence embeddings. After applying max pooling to the embeddings, the resulting vectors are combined using a Hadamard product. The joint representation is passed through a multi-layer perceptron (MLP) classification head, followed by a ReLU activation function, to produce an affinity score as the final output. The error is used to fine tune the sequence embeddings.

In addition, we employed a Bidirectional Long Short-Term Memory (BiLSTM) model (36) as our baseline. Each protein sequence was input into a BiLSTM architecture with a latent dimension of 10. The outputs from the last hidden layers served as encodings for the respective sequences. Subsequently, the encoded representations from both sequences were concatenated and fed into linear layers to produce a single output representing the predicted binding energy. To address the positional bias inherent in this architecture, where residues near the beginning of the sequence disproportionately influence predictions, we implemented an enhanced inference strategy. This approach processes protein sequences using a sliding window technique with overlapping chunks (100 residues per chunk with a 50-residue stride). Each chunk of residues is independently analyzed by the BiLSTM model. The resulting predictions are then averaged, which ensures uniform consideration of residues regardless of their position in the sequence Overall, this approach significantly improves the model's ability to identify critical interaction sites throughout the entire protein chain rather than focusing primarily on N-terminal regions.

**Evaluation**

To assess the predictive accuracy of the PLMs, a set of 14 protein pairs, relevant for human health with experimentally confirmed interactions, was selected for targeted evaluation. These include immune-related complexes, such as 4UDT (superantigen–TCR) (37), 6ISC (NK cell activation) (38), 7SO0 (chemokine neutralization) (39), 7DC7 (ATP-dependent antibody) (40), and 6WYT (antigen-binding fragments) (41), relevant in infection, inflammation, and immunotherapy. Cancer-associated interactions are represented by 6XI7 (KRAS–RAF1) (42), 8EJM (mutant spliceosome complex) (43), and 5V89 (ubiquitin ligase regulation) (44). Neurological and vascular signaling is illustrated by 5WBX (SK channels) (45), 7F1G (BACE1 inhibition in Alzheimer's) (46), and 6UZK (HEG1-KRIT1 complex in cerebrovascular disease) (47). Therapeutic and viral-host models include 6UMI (migraine antibody–receptor) (48), 7E50 (plasmin–inhibitor in fibrinolysis) (49), 6T36 (HBV–PDZ signaling) (50), 4U7E (ESCRT machinery) (51), and 7WBP (Omicron RBD–hACE2 binding) (52). Together, these structures span major biological systems implicated in human disease.

To further analyze the structures and binding interface, we conducted a secondary structure analysis of all the pdb complexes. For this, we utilized the DSSP (Define Secondary Structure of Proteins) algorithm (53). Additionally, in order to comprehensively assess the models' robustness and generalization capabilities, a series of mutations were introduced for each protein pair. Specifically, for a given protein pair, one protein sequence remained unaltered, while the sequence of the interacting partner underwent systematic mutations. These mutations involved replacing individual amino acids with alanine (A), as well as more complex alterations, such as substituting each pair of consecutive amino acids with di-alanine (AA) and extending this pattern to larger subsequences. This iterative mutation process continued until each set of 5 or 10 consecutive amino acids was replaced by 5 or 10 alanine residues. This sequence manipulation serves to simulate a diverse range of mutation scenarios, providing a comprehensive evaluation of the models' predictive capabilities under various conditions. Additionally, we utilized RING 4.0 to investigate whether the models exhibit different success rates when predicting interface residues involved in specific interaction types (54). These interactions involve hydrogen bonds, ionic interactions, π-cation, π-π stacking, and disulfide bonds.

While larger patch sizes (e.g., 5 and 10) facilitate the exploration of spatially extended mutational effects, evaluating performance based solely on predicted ddG magnitude can be challenging due to the potentially subtle changes in binding affinity induced by single-site substitutions. To address this, we established an evaluation criterion based on an "N Factor". This factor is applied as a multiplier to the size of the true interface residue set, denoted as *n*. Specifically, performance is assessed by evaluating the model's ability to rank true interface residues among the top *k* predicted positions, where *k* is defined as $N \times n$. We performed this analysis using N factors of 1, 2, and 3. For instance, for a complex with 20 true interface residues (n=20), we evaluated predictions against the top 20 (1*n), 40 (2*n), and 60 (3n) highest-scoring positions. Finally, we benchmarked the model performance against a state-of-the-art structural docking method (HADDOCK3) (55).

**RESULTS**

We first fine-tuned ProtBERT, ProtT5, ESM2 and BiLSTM on the SKEMPI 2.0 mutation dataset. The training and validation loss curves, shown in Figure 2, illustrate the efficient learning process of the four models. Specifically, ProtBERT converged within 5 epochs, while ProtT5 achieved convergence in just 2 epochs. Furthermore, the validation MAE achieved by each model, presented in Figure 3 and detailed in Supplementary File 1, demonstrates consistently low prediction errors across architectures. These results underscore the effectiveness of the fine-tuned models in capturing protein-protein binding affinities, particularly given that a majority of experimental binding scores in the SKEMPI 2.0 dataset fall within the 5–15 kcal/mol range (Figure 3). The relatively small MAE values compared to the dynamic range of binding affinities suggest that the models are well-calibrated and capable of capturing meaningful biophysical variations in protein interaction strength.

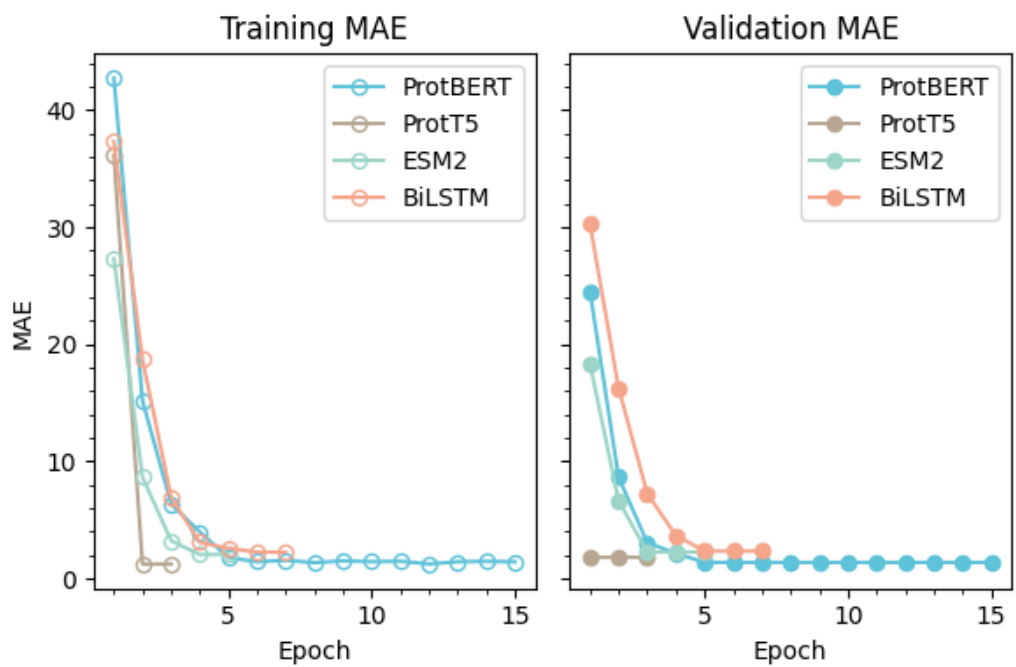

Figure 2: Loss curves for ProtBERT, ProtT5, ESM2, and BiLSTM across training epochs. Both training and validation MAE are shown using distinct marker styles. Among the models evaluated, ProtBERT achieves the lowest validation MAE, indicating the best overall predictive performance.

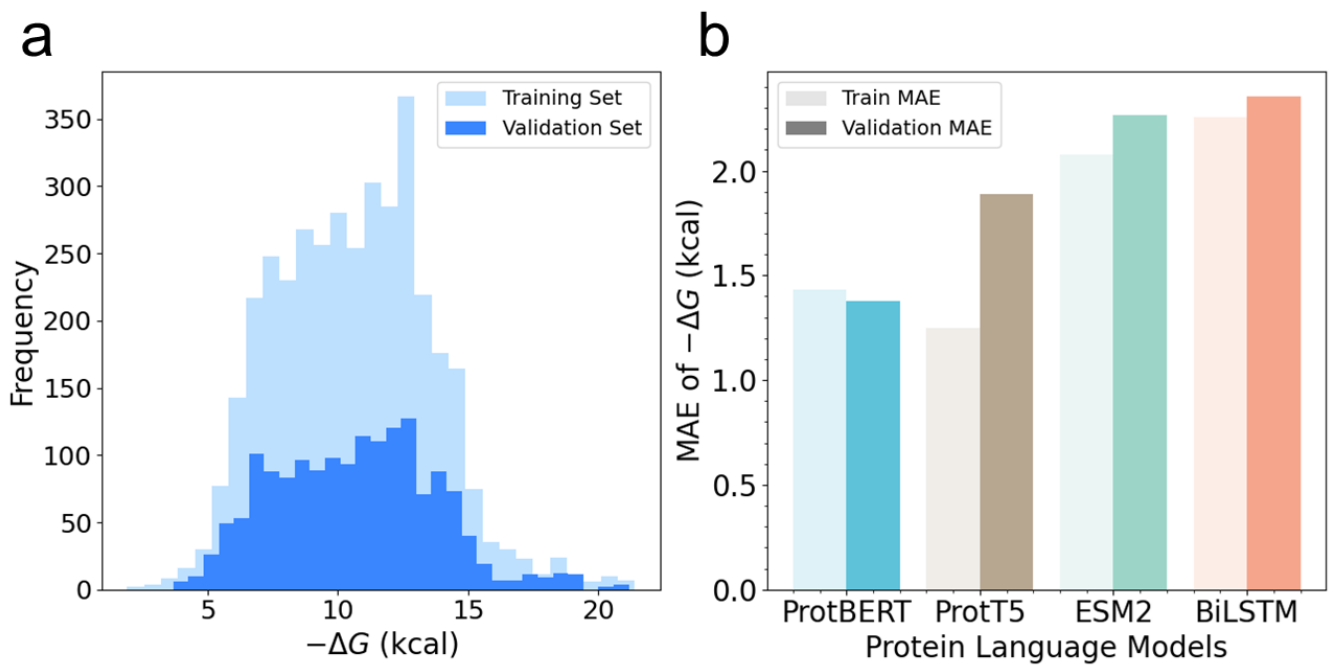

Figure 3: (a) Distribution of −ΔG values for the training and validation dataset. (b) MAE of −ΔG values for each model on training and validation datasets

To evaluate the predictive accuracy of PLMs in identifying interaction residues, we assessed the four models using a set of 14 protein pairs with experimentally confirmed interactions. One protein in each pair retained its original sequence, while the interacting partner underwent either single amino acid substitutions (1 mutation) or replacement of 5 or 10 consecutive amino acids with alanines (5 or 10 mutations). Initially, we mapped the sequences of the 14 PDB complexes. However, two complexes exhibited poor mapping quality (Supplementary File 1). Consequently, we proceeded with 14 out of the original 14 complexes for subsequent analysis.

As shown in Figure 4, model performance varies markedly across mutation scenarios. This performance was measured as the fraction of true amino acids captured among the top n predictions. For single-mutation targets, and N Factor 1, ESM2 leads the field with an accuracy of 18.0 %, followed by ProtT5 at 16.5 %, ProtBERT at 14.8 %, and our BiLSTM baseline at 11.9 %. Contrary to our initial hypothesis, enlarging the patch size offers only marginal gains in prediction quality (see Supplementary Fig. S1), suggesting that contextual breadth alone is not the primary driver of improved residue recovery.

By contrast, elevating the "N factor", the number of candidate residues considered per position, yields significant improvements. As illustrated in Figure 4, moving from an N factor of 1 to 3 nearly triples ESM2's accuracy, from 18.0 % to 49.5 %. Similar upward trends are observed for all models, underscoring that allowing multiple top-ranked hypotheses per site substantially enhances the likelihood of including the correct substitution. Moreover, this benefit is consistent across each protein complex tested, highlighting the robustness of the multi-candidate approach in capturing true mutational events (Figure 4).

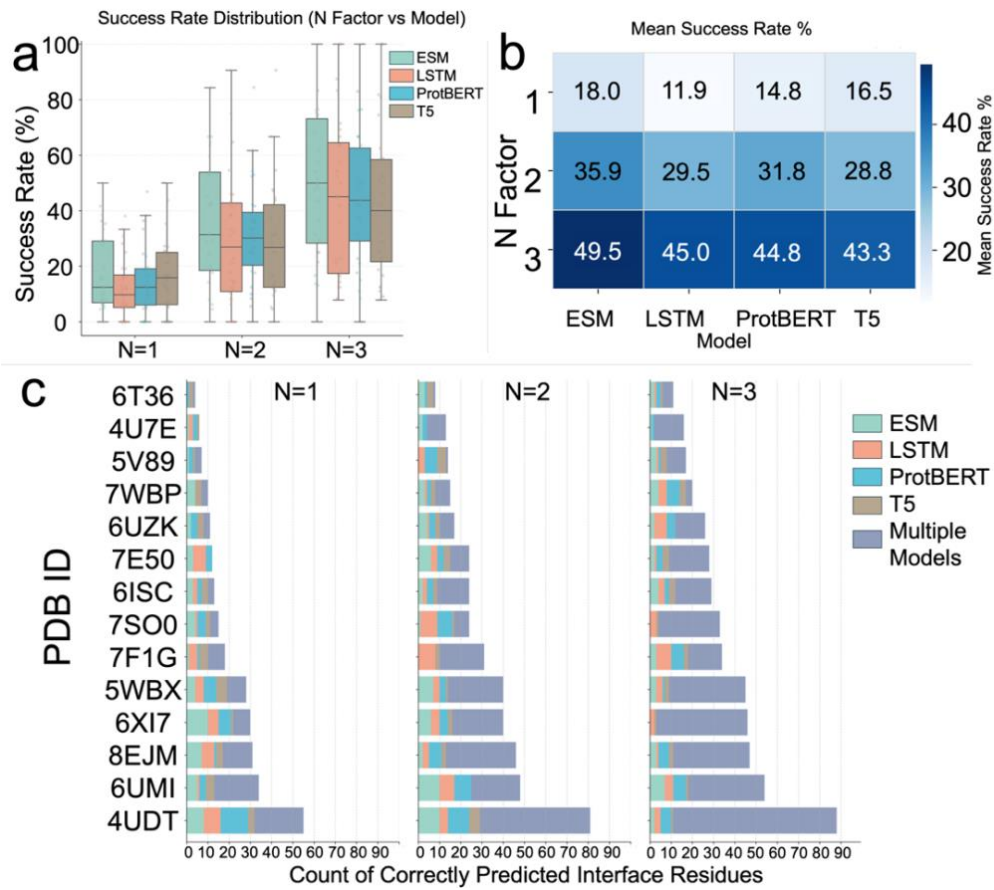

Figure 4: Performance Comparison of Interface Residue Prediction Models Across Different N factors. (a) Distribution of success rates (%) for each model (ESM2, BiLSTM, ProtBERT, T5) across analyzed protein complexes for N factor 1, 2, and 3. Success rate is defined as the average percentage of ground truth interface residues correctly predicted by the model for 14 complexes. (b) Heatmap showing the mean success rate % across different models (x axis) and N factors (y axis). (c) Stacked horizontal bar charts showing the count of correctly predicted interface residues per PDB ID for each model. The total length of the bar represents the sum of correct predictions by all models for that complex (including residues predicted correctly by multiple models). Three different parts represent results for N factor 1,2, and 3.

Figure 4 reveals significant heterogeneity in prediction performance across different protein complexes. Complexes such as 4UDT and 8EJM consistently show higher counts of correctly identified interface residues compared to others like 6UZK or 7WBP, irrespective of the patch size or N-factor size. Increasing the N factor led to a significant rise in correct predictions, notably enabling models that initially failed to identify any correct residues at stricter thresholds to successfully capture true positive sites. For instance, the LSTM model, which identified no correct residues at Top 1N (Patch 1), successfully predicted interface residues at Top 3N (Supplementary Fig. S2).

Furthermore, the relative contribution of each model (ESM2, BiLSTM, ProtBERT, T5) varies substantially depending on the specific complex. While certain models like ESM2 might dominate predictions for some PDB IDs, others like ProtT5 or ProtBERT demonstrate stronger performance for different complexes (e.g., ProtT5 in 5WBX and 8EJM) (Supplementary Fig. S3). Comparing the plots across patch sizes indicates that the total number of correct predictions and the best-performing model can be complex-dependent, suggesting no single patch size is universally optimal (Supplementary Fig. S3). These findings highlight that interface prediction difficulty is highly variable between protein systems.

Table 1: Predictive accuracy of models for N factor 1, Patch 1 for different types of residue bonds

| Bond Type | Ground-Truth Count | ESM2-Recovery% | BiLSTM-Recovery% | ProtBERT-Recovery% | T5-Recovery% |
|---|---|---|---|---|---|
| H-bond | 87 | 19.5 | 11.5 | 16.1 | 18.4 |
| Ionic bond | 6 | 33.3 | 50.0 | 16.7 | 0.0 |
| $\pi$-Cation | 5 | 0.0 | 20.0 | 80.0 | 20.0 |
| $\pi\pi$-Stacking | 15 | 26.7 | 13.3 | 26.7 | 20.0 |
| Disulfide | 2 | 0.0 | 50.0 | 50.0 | 50.0 |

To investigate if model performance is influenced by the underlying interaction chemistry, we analyzed the recovery rate for residues involved in specific bond types, focusing on Patch 1 results (Table 1). The ground truth interaction types were determined using RING 4.0 based on the reference structures. However, RING 4.0 did not identify interactions for every residue within the 6.5 Å interface definition used for the primary hotspot analysis, providing a subset of chemically characterized interactions. Focusing on specific non-covalent interactions (excluding Van der Waals (VdW)), we observed distinct recovery patterns. A key finding was the consistently low recovery rate (11.5-19.5%) for hydrogen-bonded residues across all models, despite H-bonds representing the largest category of identified non-VDW interactions (N=87). This suggests identifying H-bond partners from sequence alone presents a significant challenge for these zero-shot models. While performance varied more substantially between models for rarer interaction types (ionic, $\pi$-cation, $\pi$-$\pi$ stacking, disulfide), the low number of ground truth instances in these categories (N in the range of 2 to 15) limits robust interpretation of apparent model-specific strengths or weaknesses.

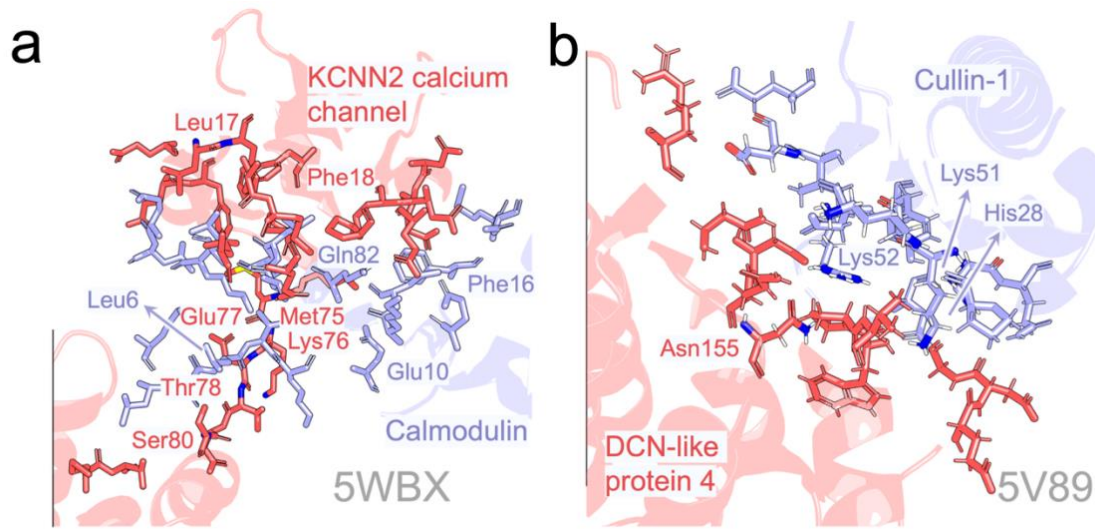

Figure 5: Interface-residue recovery on two benchmark complexes (PDB IDs 5WBX and 5V89). For each complex we display the experimentally validated interacting residues as sticks (ground truth) and annotate those positions that are recovered by at least two of the four models tested for (a) 5WBX (a SK/IK channel positive modulators) and by our best-performing model esm2 for (b) 5V89 (a DCN1-like protein 4 PONY domain bound to Cullin-1).

For complex 5WBX, most correctly retrieved contacts lie in the hydrophobic core of the interface, including Leu 6, Phe16, Leu17, Phe18, and Met 75 from both chains. Increasing the top N threshold markedly boosts coverage. With N = 1 the four models recover an average of 12 true contacts, which rises to 20 and 28 at N = 2 and N = 3, respectively. Increasing the N factor benefits ProtBERT significantly, as its hit count increases four to five-fold with the broader candidate list (Supplementary File 1), Patch-size enlargement, in contrast, confers only marginal additional gains (Supplementary Fig S4). For ProtT5, we observed a trend of decreasing performance with larger patch sizes. On the other hand, complex 5V89 is dominated by basic side chains. ESM-2 reliably recovers His28, Lys51, and Lys52, and, at higher N values, additionally identifies Arg111 on chain A and Asp55, Glu59, Glu61, Arg65, and Asp71 on chain B. Overall success rates climb sharply when moving from N=1 to N= 3 (Supplementary File 1), underscoring the value of permitting multiple ranked hypotheses per site. Together, these snapshots illustrate the model's capacity to pinpoint key interface determinants, particularly when a modest expansion of the prediction set is allowed, while also revealing complexes where simpler architectures such as the BiLSTM still underperforms.

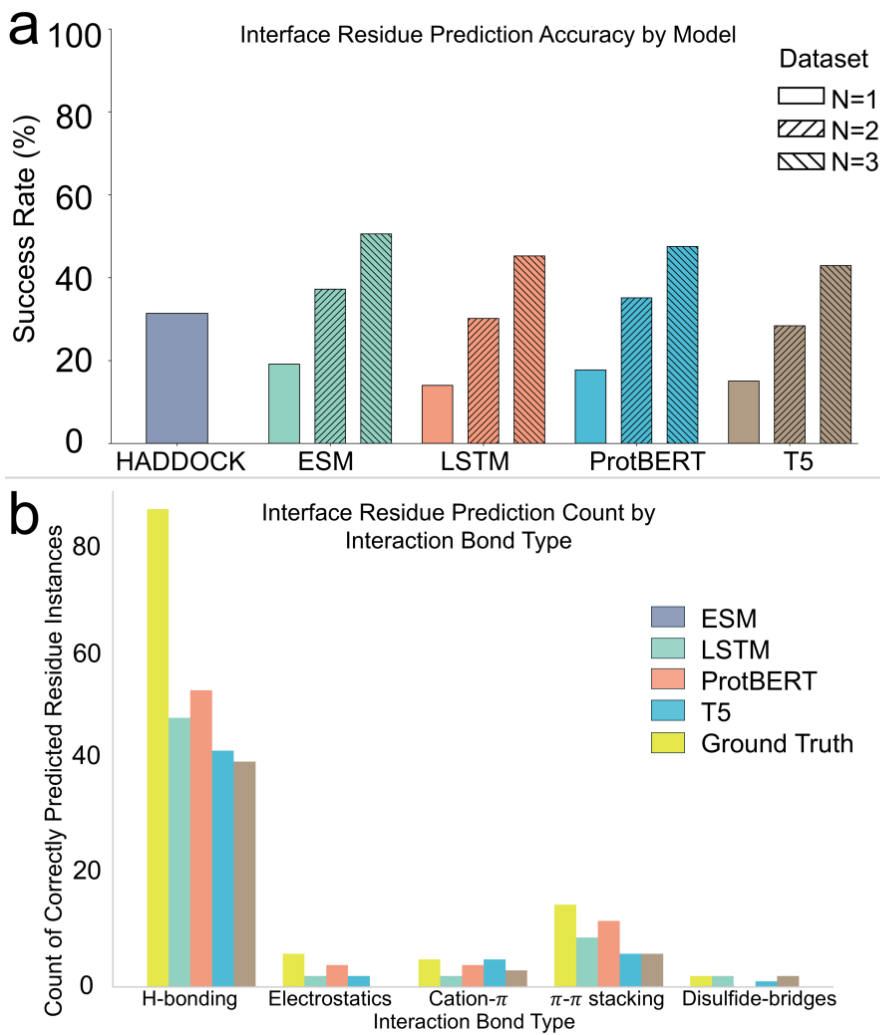

Figure 6: Performance comparison of protein interface prediction methods. (a) Success rates (%) of language model-based predictors (ESM2, BiLSTM, ProtBERT, T5) and structural docking (HADDOCK3) for interface residues across three N factors (N=1,2,3). (b) Distribution of correctly predicted interaction types across 14 protein complexes, shown as the number of ground truth interactions recovered by each method.

The comparative performance of our protein interface prediction approach is shown in Figure 6. At higher contextual depth (N-factor=3), Seq2Bind shows enhanced performance in detecting interacting residues. Even at moderate contextual depth (N-factor=2), ESM (37.2%) and ProtBERT (35.1%) outperform state-of-the-art blind docking with HADDOCK3 (32.1%).

Subsequently, the observed trend of higher accuracy in the 10-mutation scenario compared to single mutations can be attributed to how the models process sequence data and detect patterns. Transformer-based PLMs, such as ProtBERT, ProtT5, and ESM2, rely heavily on self-attention mechanisms to capture long-range dependencies and contextual information across entire sequences. In the 10-mutation case, the extensive sequence alterations introduce a more prominent signal of disruption in the protein sequence, making it easier for these models to identify significant changes. By contrast, single mutations represent subtler alterations, which may be harder for the models to discern amidst the inherent variability of natural protein sequences.

In contrast to transformer-based models, the BiLSTM, which serves as a baseline, processes sequences sequentially, leveraging recurrent mechanisms to capture temporal dependencies in both forward and backward directions. While BiLSTM does not use attention mechanisms, its ability to generalize in the 10-mutation scenario may stem from its focus on capturing local sequential patterns rather than relying on long-range relationships. The higher accuracy of BiLSTM in this scenario suggests that it may be less sensitive to overfitting subtle changes, offering a complementary perspective on handling sequence disruptions.

From an NLP perspective, PLMs treat protein sequences as tokenized data and learn statistical patterns across these tokens. Larger, contiguous mutations create a clearer anomaly relative to the learned patterns of natural sequences, making it easier for PLMs to detect interaction residues. Single mutations, however, lack this pronounced contextual disruption, posing challenges for both PLMs and BiLSTM in distinguishing biologically significant mutations from background noise.

**DISCUSSION**

Despite the growing utility of PLMs in modeling PPIs, our findings highlight critical limitations in their ability to reliably identify residue-level interaction sites through fine-tuning. Although ProtBERT, ProtT5, and ESM2, when fine-tuned on binding affinity data, exhibited strong predictive accuracy for global binding strength, their performance in pinpointing interaction interfaces across diverse protein complexes remained inconsistent. Notably, while alanine-scanning mutagenesis revealed a subset of correctly predicted interface residues, the recovery rates, particularly for hydrogen-bonded contacts, were uniformly low across all models. This suggests that single-point mutations introduce subtle sequence perturbations that PLMs struggle to associate with local structural or energetic disruptions, particularly in the absence of explicit 3D structural context. Moreover, the predictive accuracy varied markedly between complexes and interaction chemistries, suggesting that fine-tuned models remain sensitive to dataset bias and system-specific variability. The higher accuracy observed in larger sequence perturbations (e.g., 10-residue alanine patches) further suggests that PLMs may be more attuned to detecting broader contextual disruptions than fine-scale contact determinants.

To improve the interpretability and robustness of interaction residue prediction, future studies could integrate structural priors during model training, such as incorporating residue-residue distance maps or attention-guided contact constraints from experimentally resolved structures. Multi-modal training frameworks that jointly learn from both sequence and structure could bridge the gap between global sequence embeddings and local geometric features critical for interface formation. Additionally, rather than relying solely on alanine scanning, leveraging generative adversarial mutation strategies (56) or in silico evolution simulations (57) may better expose latent model biases and enhance sensitivity to biologically relevant perturbations. Benchmarking models across chemically annotated interaction types (e.g., hydrogen bonds, ionic, π-stacking) with balanced ground-truth sets will also be essential to isolate model-specific strengths and limitations. Finally, expanding training datasets to include more diverse and experimentally resolved complexes, especially those with non-canonical or transient interactions, may improve generalization and help disentangle sequence-derived signals from structure-imposed constraints.

Our analysis revealed that increasing the N Factor, and thus the number of top-ranked predictions considered, significantly improves the success rate of protein language models in identifying known interface residues (Figure 4). Consequently, model selection and threshold tuning are crucial for downstream applications. The demonstrated increase in the count of unique correct predictions with higher N Factors (Figure 4) confirms that expanding the prediction threshold effectively captures a greater number of distinct true positive interface sites, providing a more comprehensive map of the predicted interaction surface. Further insights into the characteristics of these top predictions were gained by examining their associated predicted ddG values (Supplementary Fig. S5). Consequently, since the active residues in protein-protein interactions typically remain unknown before experimental characterization, molecular docking and simulations are often unsuccessful and waste computational resources. With structural information unavailable for most of the proteome, and considering the challenges posed by disordered regions and non-globular proteins, methods that bypass structural requirements address a critical need (58)

Our analysis of the top predicted residues by the 4 models that were not part of ground truth revealed substantial spatial dispersion from the true interface region. Distance calculations between predicted residues and the nearest true interface residues showed that polar contacts predominated, suggesting electrostatic interactions play a key role (Supplementary File 1). While many predicted residues lacked direct interface contact, they may represent components of allosteric networks that influence binding through long-range effects. The prevalence of polar interactions supports this interpretation, as hydrogen bonding networks facilitate allosteric communication (Supplementary Fig. S6) These findings suggest that prediction models may capture a broader functional landscape. Overall, they identify both direct interface residues and allosterically coupled sites important for protein-protein binding.

For certain models, notably LSTM and T5, the predicted ddG values among the top N, 2N, and 3N predictions exhibited a remarkably narrow range, less than 0.05 kcal/mol for the complexes. This limited magnitude and range of predicted effects among the top candidates may suggest that these models face challenges in accurately predicting subtle changes in binding affinity resulting from mutations. This difficulty could potentially stem from inherent limitations in predicting nuanced energetic effects. Additionally, characteristics of the training data, such as potential biases or limitations in the diversity of experimental mutation data available in our SKEMPI v2 dataset, can impact the robustness of models for fine-grained ddG prediction across all mutation types and structural contexts (59). Nevertheless, the ability of these models to enrich for known interface residues through computationally predicted alanine scanning, particularly when appropriate prediction thresholds are applied via metrics like the N Factor, underscores their utility as a valuable tool for guiding experimental efforts in the characterization of protein interaction interfaces.

Overall, our approach provides a valuable preliminary step before resource-intensive molecular docking, offering superior accuracy (ESM N-factor=2 vs. HADDOCK3, Figure 6), significantly reduced computational overhead (minutes versus hours for complete docking protocols), and broader applicability to proteins lacking experimentally determined structures.

## DATA AVAILABILITY

The tool is available as a part of AgriVax.AI at https://agrivax.onrender.com under StructF suite. All relevant source code to reproduce the results in this study is publicly available via the following Google Colab link: https://colab.research.google.com/drive/1ssETy-TXgpGzcS0AhZrDvT3cOaF5pWrF, which includes both the binding energy prediction and alanine scanning mutagenesis tools. The following pretrained models were sourced from Hugging Face.

1. ESM2: https://huggingface.co/facebook/esm2_t33_650M_UR50D
2. ProtBERT: https://huggingface.co/Rostlab/prot_bert_bfd
3. ProtT5: https://huggingface.co/Rostlab/prot_t5_xl_half_uniref50-enc

## SUPPLEMENTARY DATA

Supplementary Data are available at NAR Online.

## AUTHOR CONTRIBUTIONS

RC conceived the idea and designed the project. QL and MX performed fine tuning of the language models. SD performed training data standardization, model performance analyses, and all benchmarking with experimental data. VSR performed the docking analysis and comparison with Seq2Bind. CZ helped in model training and hyper parameter optimization. All authors contributed to the writing and editing of this manuscript.

## FUNDING

This work is partially supported by the National Science Foundation under Grant No. 2152117 to QL, and by the Iowa State University Startup Grant (Building a World of Difference Faculty Fellow), Center for Industrial Research and Service (CIRAS) Mini Grant, and NSF 22-599, Established Program to Stimulate Competitive Research (EPSCoR) RII Track-1, Award Number DQDBM7FGJPC5 to RC. Any opinions, findings, conclusions or recommendations expressed in this material are those of the author(s) and do not necessarily reflect the views of the National Science Foundation or CIRAS.